\documentstyle[prb,epsf,aps,multicol]{revtex}
\draft 

\def\il{I_{low}} 
\def\iu{I_{up}} 
\def\eeq{\end{equation}}

\def\prb{Phys. Rev. {\bf B}}
\def\pra{Phys. Rev. {\bf A}} 
\def\prl{Phys. Rev. Lett. }  
\def\ajp{Am. J. Phys. }  
\def\mpl{Mod. Phys. Lett. {\bf B}} 
\def\ijmp{Int. J. Mod. Phys. {\bf B}}

\def\pjp{Pramana J. Phys.}
\begin{document}

\draft

\title{Role of quantum entanglement due to a magnetic impurity on
current magnification effect in mesoscopic open rings}

\author{Colin Benjamin\cite{coline}, Sandeep K. Joshi\cite{joshie},
Debendranath Sahoo\cite{sahooe} and A. M. Jayannavar\cite{amje} }

\address{Institute of Physics, Sachivalaya Marg, Bhubaneswar 751 005,
Orissa, India}

\date{\today}
 
\maketitle 

\begin{abstract}
  
We study the current magnification effect in presence of exchange
scattering of electron from a magnetic impurity placed in one arm of
an open mesoscopic ring. The exchange interaction causes
entanglement of electron spin and impurity spin. Earlier studies
have shown that such an entanglement causes reduction or loss of
interference in the Aharonov-Bohm oscillations leading to
decoherence. We find however, that this entanglement, in
contradiction to the naive expectation of a reduction of current
magnification, leads to enhancement as well as suppression of the
effect. We also observe additional novel features like new
resonances and current reversals.

\end{abstract}
\vskip 1.0in
\pacs{PACS Nos.: 73.23.-b, 5.60.Gg, 72.10.Bg, 72.25.-b }
\begin{multicols}{2}
  
Research on transport in mesoscopic open rings has provided several
counterintuitive results
\cite{imry_book,datta,psd,webb_ap,gia,wgtr,deo_cm,pareek_cm,deo_mpl,bcgupta}.
The classic example of one such result is the observation of normal
state Aharonov-Bohm (AB) effect\cite{webb_ap}. Among other such
effects in the context of mesoscopic systems, attention may be drawn
upon the current amplification effect in the open ring
system\cite{deo_cm,pareek_cm} wherein a current flowing in the lead is
amplified, after passage through the junction, in one of the ring arms
and appears as a negative current in the other ring arm. Like AB
effect, this effect too is purely quantum mechanical in origin. This
effect has been extended to thermal currents\cite{mosk} and to spin
currents in the presence of Aharonov-Casher flux\cite{choi}.
  
Recently we have studied the effect of decoherence on AB
oscillations in mesoscopic open ring\cite{joshi}. In the simple model that we
consider, a magnetic impurity in one arm of the ring is coupled via
exchange interaction to the incoming electron. Though the
interaction induces spin-flip scattering there is no exchange of
energy.  This fact along with entanglement\cite{schul} shows that spin-flip
scattering arising due to exchange coupling with magnetic impurity
reduces the AB oscillations and leads to decoherence\cite{joshi}.
The study of decoherence is important to understand the evolution of
quantum system towards a classical one (destruction of entanglement
due to environment). It is worthwhile to note that coupling to a
single impurity or single degree of freedom is enough to suppress
the quantum effect\cite{imry_book,sai}. In our case the
impurity does not dynamics of its own. Generally, it is believed
that only coupling with infinite degrees of freedom (a thermal bath)
could induce dephasing.  We know that the current magnification
effect, to be discussed below, is a quantum effect.  Hence one would
like to study whether quantum entanglement arising due to
spin-impurity coupling can reduce this effect. The answer seems to
be no, and the effect depends sensitively on the details of the
geometry and other physical parameters of the problem.
  
In the case of a mesoscopic loop with unequal arms connected to two
electron reservoirs at chemical potentials $\mu_1$ and $\mu_2$ via
ideal leads currents $\il$ and $\iu$ flow in the lower and upper arm
of the loop such that total current $I=\il+\iu$ is conserved in
accordance with Kirchoff's law. In general these two currents differ
in magnitude and are individually smaller than the total current
$I$. However, in certain range of Fermi energies the current $\il$
or $\iu$ may become larger than the total current $I$. This is the
current magnification effect. To conserve the total current at the
junctions, the current in the other arm becomes negative i.e., flows
against the applied external field. This negative current continues
to flow in the loop as a circulating current. The magnitude of the
circulating current is the same as that of the negative current. The
circulating current here arises in the absence of magnetic field.
The effect of a simple impurity (an impurity without any internal
structure) in one arm of the ring on the current magnification
property has been studied earlier\cite{pareek_cm}. It was found that
the impurity could not only suppress but also enhance the current
magnification depending on the Fermi energy. Thus the impurity plays 
a dual role as far as current magnification is concerned. The
experimental possibility of the observation of this current magnification
effect by looking at the orbital magnetic response of the loop is
mentioned in Refs. \onlinecite{deo_cm,pareek_cm}.
  
In the following we study the current magnification effect in
presence of a magnetic impurity in one arm of the ring. Consider a
single channel loop of circumference $L$ coupled to two electron
reservoirs by two ideal leads (see Fig. 1). We introduce a magnetic
impurity atom (referred to as spin-flipper, or the flipper, for
short) in one arm (upper) of the ring at length $l_3$ from junction
J1 and $l_4$ from J2. The lower arm is of length $l_2$ and
$L=l_2+l_3+l_4$. The spin of the electron ($\vec{\sigma}$) is thus
coupled to the spin of the flipper ($\vec S$) via the exchange
interaction $-J \vec{\sigma} \cdot \vec{S} \delta(x-l_3)$.  This
leads to scattering of the electron in which both the spin states of
the electron and the impurity could change without any exchange of
energy. This interaction conserves the total spin
($\vec{\sigma}+\vec{S}$) as well as the $z$-component of the total
spin. The two reservoirs are kept at chemical potentials $\mu _1$
and $\mu _2$ respectively. When $\mu _1$ is greater than $\mu _2$, a
net current flows from reservoir 1 on the left to the reservoir 2 on
the right. Details of the ring geometry with the impurity are
indicated in Fig.~1.

We follow the standard quantum waveguide
theory\cite{wgtr,deo_cm,pareek_cm,xia}to study this problem. The
wavefunctions ($\psi 's$) for the individual segments, as indicated in
Fig.~1, can be written as below

\begin{eqnarray}
\label{wf}
\psi _1&=&(e^{ikx}+r_u e^{-ikx})\chi _m\alpha+\nonumber\\
             &&r_d e^{-ikx}\chi _{m+1}\beta,\nonumber\\
\psi _2&=&(A_u e^{ikx}+B_u e^{-ikx})\chi _m\alpha+\nonumber\\
             &&(A_d e^{ikx}+B_d e^{-ikx})\chi _{m+1}\beta,\nonumber\\
\psi _3&=&(C_u e^{ikx}+D_u e^{-ikx})\chi _m\alpha+\nonumber\\
             &&(C_d e^{ikx}+D_d e^{-ikx})\chi _{m+1}\beta,\nonumber\\
\psi _4&=&(E_u e^{ikx}+F_u e^{-ikx})\chi _m\alpha+\nonumber\\
             &&(E_d e^{ikx}+F_d e^{-ikx})\chi _{m+1}\beta,\nonumber\\
\psi _5&=&t_u e^{ikx}\chi _m\alpha+t_d e^{ikx}\chi _{m+1}\beta.
\end{eqnarray}

\noindent where, $k$ is the wave-vector of incident electron. The
subscripts $u$ and $d$ represent ``up'' and ``down'' spin states of the
electron with the corresponding spinors $\alpha$ and $\beta$ respectively
(i.e., $\sigma _z\alpha =\frac{1}{2}\alpha$, $\sigma _z\beta
=-\frac{1}{2}\beta$) and $\chi _m$ denotes the wave function of the
impurity \cite{ajp} with $S_z=m$ (i.e., $S_z\chi _m=m\chi _m$). The
wavefunctions in equation (\ref{wf}) is a correlated function of the
electron and impurity spins which takes into account the fact that the
exchange interaction conserves the $z$-component of the total spin
($\vec{\sigma}+\vec{S}$). The incident electron is taken to be spin-up
polarized. The reflected (transmitted) waves have amplitudes $r_u$ ($t_u$)
and $r_d$ ($t_d$) corresponding to the ``up'' and ``down'' spin components
respectively. Continuity of the wave functions and the current
conservation\cite{wgtr,xia,ajp} at the junctions J1 and J2 imply the
following boundary conditions.

\begin{eqnarray}
  \label{bc}
  \psi _1(x=0)=\psi _2(x=0)=\psi _3(x=0),\nonumber\\
  \psi _1^\prime (x=0)=\psi _2^\prime (x=0)+\psi _3^\prime (x=0),\nonumber\\
  \psi _4^\prime (x=l_3)-\psi _3^\prime
          (x=l_3)=G(\vec{\sigma}\cdot\vec{S})\psi _3(x=l_3),\nonumber\\
  \psi _3(x=l_3)=\psi_4(x=l_3),\nonumber\\
  \psi _4(x=l_3+l_4)=\psi _5(x=0)=\psi _2(x=l_2),\nonumber\\
  \psi _2^\prime (x=l_2)+\psi _4^\prime (x=l_3+l_4)=\psi _5^\prime (x=0).
\end{eqnarray}

\noindent 
Here $G=2mJ/\hbar ^2$ is the dimensionless coupling constant
indicative of the ``strength'' of the spin-exchange interaction.  The
primes denote the spatial derivatives of the wave functions. Equations
(\ref{wf}) along with the boundary conditions (\ref{bc}) were
solved to obtain the amplitudes, with these we evaluate the current
densities in each arm of the ring. We have taken the flipper to be a
spin-half object ($M=2$) situated symmetrically at the center of the
upper arm, i.e., we consider the case in which $l_3=l_4$. Now,
depending upon the initial state of the flipper we have possibility of
either spin-flip scattering ($\sigma _z=1/2$, $S_z=-1/2$) or no
spin-flip scattering ($\sigma _z=1/2$, $S_z=1/2$), as demanded by the
conservation of the total spin and its $z$-component. In the former
case the problem reduces to that of a simple potential scattering from
the impurity as studied earlier\cite{pareek_cm}.

We have set $\hbar=m=1$ and the value of interaction strength $G$ is
given in dimensionless units throughout. The dimensionless current
density $I_2$ in the small energy interval $dE$ (see Refs.
\onlinecite{deo_cm,pareek_cm}) flowing in lower arm of the ring is
defined as $I_2=I_{2u}+I_{2d}$, wherein $I_{2u}={\mid
A_{u}\mid}^2-{\mid B_{u}\mid}^2$ and $I_{2d}={\mid
A_{d}\mid}^2-{\mid B_{d}\mid}^2$.  The current densities $I_3$ and $I_4$
flowing in the upper arm are similarly defined. Current conservation at
the impurity site demands $I_{3}=I_{4}$ which we have verified. The total
current flowing through the arms can be calculated by integrating the
corresponding current densities over the Fermi energy interval $\mu_1$ to
$\mu_2$\cite{deo_cm,pareek_cm}. In the present work we confine ourselves
to the effect of the flipper on the circulating current density $I_{c}$.
We have studied the behaviour of the current densities $\il=I_{2}$ and
$\iu=I_{3}$ in the lower and the upper arms of the ring, as a function of
$kL$. In any range of Fermi energy, if any one of these is negative, the
magnitude of the negative current density can be identified with that of
the circulating current density\cite{deo_cm,pareek_cm} $I_c$. When $\il$
is negative the direction of the circulating current density is clockwise
and when $\iu$ is negative then it is anti-clockwise. A clockwise
circulating current density is taken to be positive and an anti-clockwise
one, negative according to the usual convention.  It has been seen earlier
that there is an enhancement in current magnification in presence of a
simple impurity(no spin-flip)\cite{pareek_cm}.  In the following
paragraphs we describe our results graphically.

Figure~2 shows the plot of circulating current density ($I_{c}$)
versus $kL$ for the two separate cases of spin-flip scattering and
no-spin-flip scattering. When the impurity spin is ``up'' the
interaction does not allow spin-flip for a spin-up incident electron
due to conservation laws mentioned above. For this case problem
reduces to that of simple impurity studied earlier as mentioned above.
On the other hand when the impurity spin is ``down'' a spin-flip
scattering takes place. We compare the circulating current densities
for these two cases in order to see the role of entanglement induced
by the spin-flipper. Given the fact that the entanglement reduces
AB-oscillations we study whether it also reduces the current
magnification. The solid curve is for the no-flip case while the
dashed one is for the spin-flip case with all other parameters kept
identical. The impurity strength ($G$) for both the cases is $4.0$.
In both the cases we take $l_{2}/L=0.6$ and $l_{3}/L=l_{4}/L=0.2$. The figure
shows that , the circulating current for
spin-flip case is significantly less than that of the no-flip case in
the range $12<kL<16$.  Thus one is led to believe that the flipper
acting as a dephasor suppresses the quantum phenomena of current
magnification.

However, this naive expectation turns out to be incorrect. This is
substantiated in Fig.~3 which shows circulating current densities for
the spin-flip and no-flip cases in the range $16<kL<19$ for the same
lengths as mentioned above. From this figure we see that in this
range of Fermi energies the amplitude of the circulating current is
actually enhanced in spite of the spin-flip scattering.

Thus the flipper can not only suppress the current magnification
effect but {\em can also} enhance it in some other range of Fermi
energies. This effect can be ascribed to multiple reflections induced
by the flipper and the junctions. The fact that entanglement due to
exchange interaction between the electron and the flipper does not
eliminate multiple reflections due to scattering at the impurity (and
at the junctions) seems to be the reason behind the aforesaid
features\cite{pareek_cm}. Thus far we have discussed how the flipper
affects well known features of the impurity current magnification
effect. However, the flipper induces some new features which we
discuss below.  The plot of circulating current density ($I_{c}$)
versus $kL$ for $l_{3}/L=l_{4}/L=0.25$ and $l_{2}/L=0.5$ in the range
$5.6<kL<6.6$ shows an additional peak in the circulating current
density arising at a point corresponding to a minimum of spin-up
transmission (which is same as the maximum of the spin-down
transmission). This is indicative of the spin-flip process. This
effect is unique for the flipper having no counterpart in case of a
simple impurity, i.e., in this region ($5.6<kL<6.2$) no-flip
scattering case does not show any circulating current. This can be
ascribed to the additional phase shifts caused by spin-flip
scattering along-with multiple reflections. In the range $6.2<kL<6.6$
spin-flip scattering suppresses the current magnification.

Further, we see yet another interesting feature, namely the phenomenon of
current reversal. This is depicted in Fig.5. In this figure we plot the
circulating current density ($I_{c}$) versus $kL$ for $l_{3}/L = l_{4}/L =
0.3125$ and $l_{2}/L=0.375$ in the wave vector range $10<kL<15$ in which we
see that the spin-flip circulating current reverses its direction as
compared to the no-flip case, i.e., an anti-clockwise circulating current
for the no-flip case is converted into a clockwise one in the spin-flip
case.

In conclusion we have shown that presence of the spin-flipper which
suppresses the AB oscillations, need not suppress the amplitude of current
magnification. In fact, in certain range of Fermi energies the flipper
enhances the current magnification. Apart from this it leads to current
reversals and current magnification in additional Fermi energy ranges. We
believe that non-suppression of this quantum effect is peculiar to elastic
scattering in presence of exchange interaction leading to entanglement
between the different degrees of freedom of the electron (space and spin)
and the impurity spin state. This model can also be extended to study the
phenomenon of quantum erasure with appropriate modifications.  Only the
presence of inelastic scattering, leading to irreversible loss of phase
memory, can dephase AB oscillations and reduce current magnification
simultaneously.
 
\acknowledgments 

One of us (DS) would like to thank Professor S. N. Behera for extending
hospitality at the Institute of Physics, Bhubaneswar.

\begin{figure}
\protect\centerline{\epsfxsize=3.5in \epsfbox{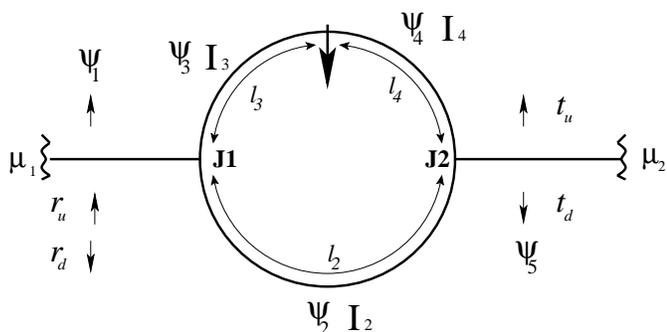}}
\caption{Mesoscopic ring with a magnetic impurity in one arm of
the ring. The situation depicted corresponds to a spin-flip
scattering process.}
\label{ring}
\end{figure}

\begin{figure}
\protect\centerline{\epsfxsize=3.5in \epsfbox{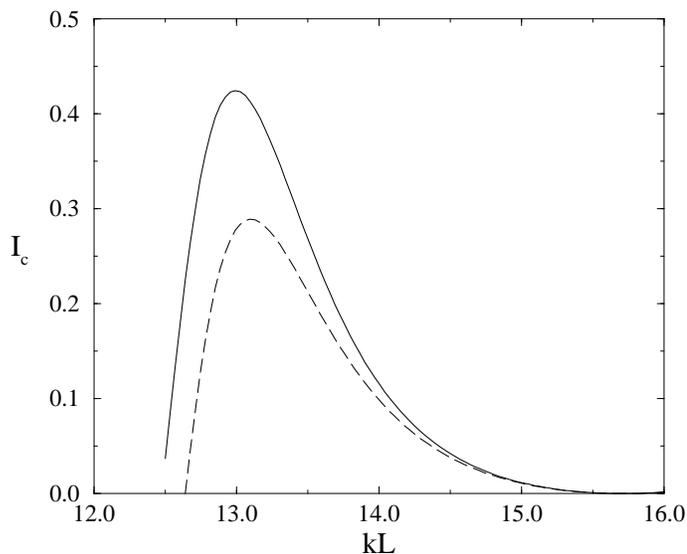}}
\caption{Plot of circulating current density $I_{c}$ versus $kL$. $G=4.0$ and
$ l_{2}/L = 0.6, l_{3}/L = l_{4}/L = 0.2 $ for both cases. 
The solid line is for the
no-flip case while the dashed line is for the spin-flip case.This
figure shows that the spin-flip process inhibits current
magnification.}
\label{fig1}
\end{figure}
 
\begin{figure}
\protect\centerline{\epsfxsize=3.5in \epsfbox{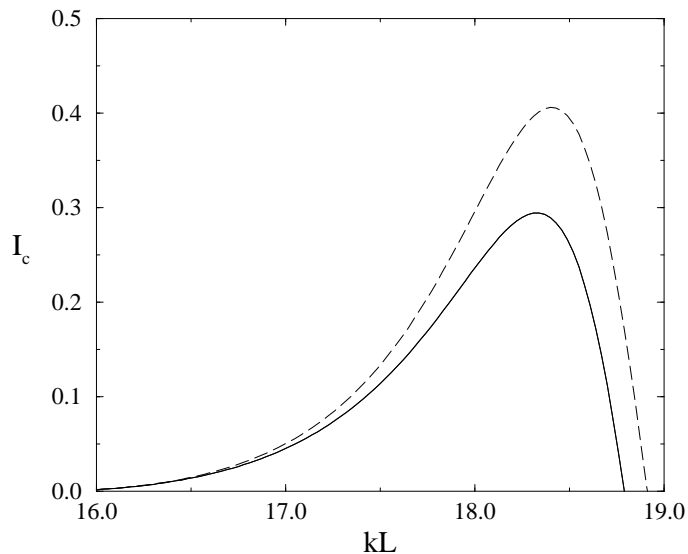}}
\caption{Plot of circulating current density $I_{c}$ versus $kL$. $G=4.0$ and
$ l_{2}/L = 0.6, l_{3}/L = l_{4}/L = 0.2 $ for both cases. 
The solid line is for the
no-flip case while the dashed line is for the spin-flip case. This
figure in contrast to Fig. 2 shows that the spin-flip process
enhances current magnification.}
\label{fig2}
\end{figure}
 
\begin{figure}
\protect\centerline{\epsfxsize=3.5in \epsfbox{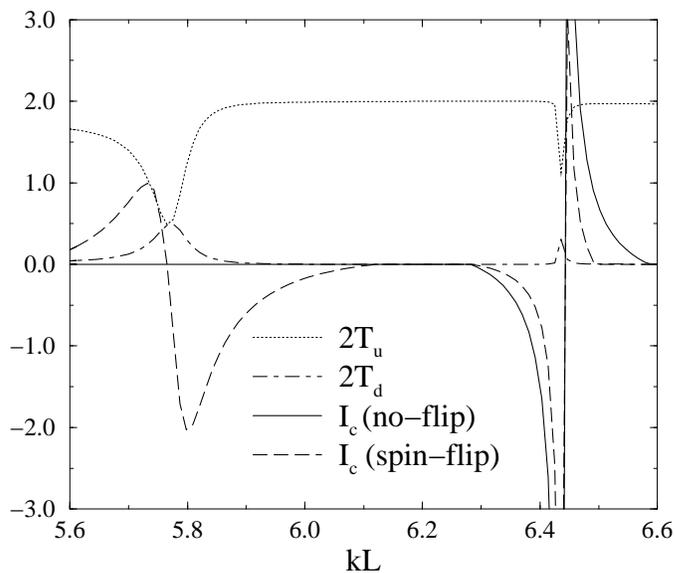}}
\caption{Plot of circulating current density $I_{c}$ versus $kL$. $G=4.0$ and
$ l_{2}/L = 0.5, l_{3}/L = l_{4}/L = 0.25 $ for both cases. The solid line
is for the no-flip case while the dashed line is for the spin-flip
case.The dash-dotted line is for $2T_{d}$ while the dotted line is
for $2T_{u}$ wherein $T_{u} = {\mid t_{u}\mid}^2$ and $T_{d} = {\mid
t_{d}\mid}^2$.}
\label{preso}
\end{figure}

\begin{figure}
\protect\centerline{\epsfxsize=3.5in \epsfbox{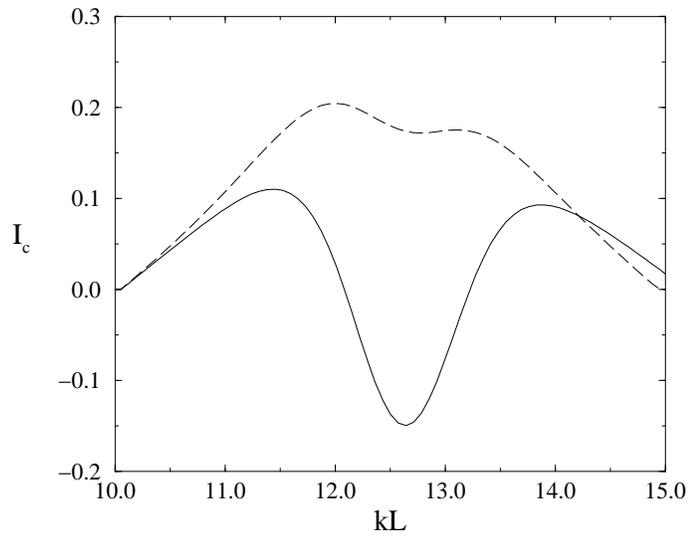}}
\caption{Plot of circulating current density $I_{c}$ versus $kL$. $G=4.0$ and
$l_{2}/L = 0.375, l_{3}/L = l_{4}/L = 0.3125$ for both cases. The solid
line is for the no-flip case while the dashed line is for the
spin-flip case.}
\label{prev}
\end{figure}
\end{multicols}
 
\end{document}